# Measuring Phonon Dispersion and Electron-Phason Coupling in Twisted Bilayer Graphene with a Cryogenic Quantum Twisting Microscope

J. Birkbeck[1†], J. Xiao[1†], A. Inbar[1†], T. Taniguchi[2], K. Watanabe[2],

E. Berg[1], L. Glazman[3], F. Guinea[4,5], F. von Oppen[6] and S. Ilani[1*]

[1] *Department of Condensed Matter Physics, Weizmann Institute of Science, Rehovot 76100, Israel.*
[2] *National Institute for Materials Science, 1-1 Namiki, Tsukuba, 305-0044 Japan.*
[3] *Department of Physics, Yale University, New Haven, CT, USA.*
[4] *Imdea Nanoscience, Faraday 9, 28049, Madrid, Spain.*
[5] *Donostia International Physics Center, Paseo Manuel de Lardizábal 4, 20018 San Sebastián, Spain.*
[6] *Dahlem Center for Complex Quantum Systems and Fachbereich Physik, Freie Universität Berlin, 14195 Berlin, Germany.*
† These authors contributed equally to the work.
* Correspondence to: shahal.ilani@weizmann.ac.il

**The coupling between electrons and phonons is one of the fundamental interactions in solids, underpinning a wide range of phenomena such as resistivity, heat conductivity, and superconductivity. However, direct measurements of this coupling for individual phonon modes remains a significant challenge. In this work, we introduce a novel technique for mapping phonon dispersions and electron phonon coupling (EPC) in van der Waals materials. By generalizing the quantum twisting microscope[1] to cryogenic temperatures, we demonstrate its capability to map not only electronic dispersions via elastic momentum-conserving tunnelling, but also phononic dispersions through inelastic momentum-conserving tunnelling. Crucially, the inelastic tunnelling strength provides a direct and quantitative measure of the momentum and mode resolved EPC. We use this technique to measure the phonon spectrum and EPC of twisted bilayer graphene (TBG). Surprisingly, we find that unlike standard acoustic phonons, whose coupling to electrons diminishes as their momentum goes to zero, TBG exhibits a low energy mode whose coupling increases with decreasing twist angle. We show that this unusual coupling arises from the modulation of the inter-layer tunnelling by a layer-antisymmetric "phason" mode of the moiré system. The technique demonstrated here opens the way for probing a large variety of other neutral collective modes that couple to electronic tunnelling, including plasmons[2], magnons[3] and spinons[4] in quantum materials.**

Electron-phonon coupling (EPC) plays a key role in determining the thermal and electrical properties of quantum materials. In monolayer graphene, for instance, an exceptionally weak EPC[5] results in ultra-high electronic mobility, micrometer-scale ballistic transport[6] and hydrodynamic behavior[7]. In contrast, the nature of EPC in moiré systems is much less understood. Various theories have attributed superconductivity[8] and the "strange metal" behavior[9,10] in magic-angle twisted bilayer graphene (MATBG) to a strong coupling of electrons to optical[11–13] or acoustic[12,14–18] phonons. Specifically, it has been emphasized that in addition to the phonons of the individual layers, twisted interfaces with quasiperiodic structures exhibit unique phononic modes involving an anti-symmetric motion of atoms in the two layers. These modes, dubbed moiré phonons[19–21] or phasons[21–24] resemble acoustic modes and constitute a new set of low-energy excitations. It was proposed that phason modes may induce strong electronic effects because the moiré pattern acts as an amplifier[16] – small shifts on the atomic scale lead to significant distortions of the moiré pattern, which in turn strongly couples to the moiré energy bands.

Existing techniques for probing phonon dispersions and EPC rely on inelastic scattering of photons (ARPES[25,26], Raman[27–29] and X-Ray[30,31]), electrons (EELS[32,33]), or neutrons[34] as well as on indirect measurement through the effect of EPC on electrical resistance[35]. However, it remains challenging to extract the EPC quantitatively, especially for the low-energy acoustic modes, which are central to the physics at low temperatures.

In this work, we developed a cryogenic quantum twisting microscope (QTM) and used it to directly map the phonon spectrum and mode-resolved EPC in twisted bilayer graphene (TBG). Our measurements of the conductance between continuously twistable graphene layers reveal a series of steps indicative of inelastic tunnelling processes involving phonon emission. The evolution of these steps with twist angle demonstrates that these processes are momentum resolved, allowing us to generate a full map of the phonon dispersion. By controlling the Fermi levels of the graphene layers and their contact interface area, we precisely determine the mode- and momentum-dependent EPC. Interestingly, we observe a low energy mode whose coupling increases with decreasing twist angle. We show that this results from a layer-antisymmetric phason mode that couples directly and strongly to the inter-layer tunnelling and discuss its significance for the low-temperature physics of TBG.

**Measuring phonon dispersion with the cryogenic QTM**

Our cryogenic QTM (Fig. 1a, methods section) consists of a custom-built cryogenic atomic force microscope (AFM) with translational and rotational degrees of freedom that allows forming a twistable two-dimensional interface between two van der Waals (vdW) heterostructures - one on its tip and one on a flat substrate. The interface is created at low temperatures, gets self-cleaned by the attractive vdW forces acting between the two heterostructures, and remains in contact throughout the experiment, including during twisting or scanning operations.

We begin our experiments with a twisted interface formed between two graphite layers, both several tens of nm thick (Fig. 1b). We set the bias across the interface, $V_b$, and measure the tunnelling current, $I$, and conductance, $G = dI/dV_b$. Fig. 1c shows a measurement of $G$ vs. twist angle, $\theta$. At zero bias (black), $G$ exhibits a pronounced peak at $\theta = 0°$ (limited by contact resistance, $R_c \sim 500\Omega$) and smaller peaks at commensurate angles of $\theta = 21.8°$ and $\theta = 38.2°$. These peaks[36,37] correspond to elastic momentum-resolved tunnelling due to the overlap of the Fermi surfaces on both sides of the interface[1,38]. At $\theta = 0°$, this is due to the alignment of the Dirac points at the corners of the first Brillouin zone (BZ), and at $\theta = 21.8°, 38.2°$ the overlap occurs at higher BZs. Away from commensurate angles, the Fermi surfaces do not overlap, and the zero bias conductance reflects momentum-non-conserving processes in the experiments, which may originate from atomic defects or tip edges. Remarkably, the measured $G$ at $\theta = 30°$ is about six orders of magnitude lower than at $\theta = 0°$, demonstrating extreme level of momentum conservation to a few parts per million. At finite bias ($V_b = 50mV$, blue), $G$ increases significantly across all twist angles, indicating the activation of inelastic tunnelling channels. The bias dependence of $G$ at $\theta = 30°$ (Fig. 1d) shows that it rises in steps, signifying the onset of discrete inelastic tunnelling processes. Similar steps were observed in scanning tunnelling microscopy studies of graphene[39] and with fixed-angle tunnelling devices[40–42] and were associated with inelastic electron tunnelling mediated by phonon emission.

Given that the QTM is a momentum-resolved tunnelling probe, it is natural to ask whether the observed inelastic processes are momentum-conserving by probing their evolution with twist angle. Fig. 1e shows $G$ measured over a wide range of twist angles ($\theta$ = 0°-55°) and as a function of $V_b$, revealing again that $G$ increases in steps with bias. Importantly, we observe that the turn-on bias of these steps varies with $\theta$. This becomes even more apparent when plotting the second derivative, $d^2I/dV_b^2$ (Fig. 1f), which exhibits sharp peaks at the steps' turn-on biases. The measurement reveals a rich spectrum of low-energy peaks that slowly disperse with $\theta$. The peaks exhibit a mirror symmetry about $\theta$ = 30°, as well as a mirror symmetry between positive and negative biases. Near $\theta$ = 21.8° and $\theta$ = 38.2°, we observe additional peaks which disperse more rapidly with $\theta$. Recalling that $\theta$ and $V_b$ are proxies for momentum and energy and that near $\theta$ = 21.8° and $\theta$ = 38.2° the tunnelling is elastic and is probing the electronic bands that disperses with the Fermi velocity[1], we recognize that the observed low energy peaks reflect modes that are significantly slower than the electronic Fermi velocity.

To understand these observations, it is instructive to turn to momentum space (Fig. 1g). Although graphite's band structure is rather complex, it is easy to see that if two graphite flakes are twisted with respect to each other by more than a few degrees, the momentum mismatch between their bands is too large to allow elastic momentum-conserving tunnelling. However, phonons have much shallower dispersion, thus phonon emission at a small $V_b$ may suffice to supply the missing momentum, $q_{ph} = 2K_D \sin(\theta/2)$ (Fig. 1h, $K_D$ is the Dirac point momentum) in an inelastic tunneling process across the interface. Phonon emission turns on when the bias is larger than the phonon energy at this momentum, $eV_b > \hbar\omega_{ph}(q_{ph})$. Thus, the positions of the $d^2I/dV_b^2$ peaks in the $\theta - V_b$ plane directly trace the phonon spectrum. Indeed, overlaying the theoretically calculated spectrum for bulk graphite[43] (dashed lines) over the experiment shows excellent agreement. Specifically, we can identify various acoustic (TA, ZA) and optical (LO, TO, ZO) branches and follow their dispersion across a wide momentum range, spanning a substantial portion of the BZ. The observed mirror symmetry about $\theta$ = 30° can be understood from the fact that for each electron tunnelling from the $K$ point in the top layer and emitting a phonon corresponding to $\theta$, there is an equivalent process with an electron tunnelling from the $K'$ point and emitting a phonon corresponding to $(60° − θ)$. However, since graphite has a complex band structure, and the twisted interface between graphite flakes has phonons both in the graphite's bulk and at the interface, it is instructive to switch to a simpler system that will allow us to better understand the underlying electron-phonon coupling mechanisms.

**Electron-Phonon coupling in Twisted Bilayer Graphene**

We turn to the main system in this paper - twisted bilayer graphene (TBG) - in which electron phonon coupling is believed to be of prime importance[9–24,26,29,44–46]. We create a tunable TBG system by bringing an hBN-backed monolayer graphene, placed on a tip, into contact with another monolayer graphene on a bottom sample that incorporates a buried graphite gate (Fig. 2a). When in contact, the Dirac cones of the two graphene layers, now at the corners of a mini-BZ, hybridize to yield the TBG energy bands (color-coded in Fig. 2c by layer weight). The $G$ and $d^2I/dV_b^2$ in this twistable TBG, measured as a function of $θ$ and $V_b$, are shown in Figs. 2d and 2e. The steps in $G$ (peaks in $d^2I/dV_b^2$) disperse with $V_b$, demonstrating that the phonon spectrum in TBG, mapped by this measurement, is similar to that in bulk graphite (dashed lines).

A key feature of our momentum-resolved inelastic tunnelling technique is that it allows us to directly determine the mode- and momentum-dependent EPC. As will be shown below, the height of the step in $G$, or equivalently the area under the peak in $d^2I/dV_b^2$ is directly proportional to the strength of the EPC. Looking at the measurements in Figs. 2d and 2e, we recognize that the EPC varies significantly between the different phonon branches. Around $V_b \sim 160mV$, we see a pronounced step in $G$ that corresponds to the optical phonons (LO and TO). Indeed, strong EPC to optical phonons is expected in monolayer graphene[47,48] and is believed to play an important role in MATBG[13,26]. More puzzling behaviors are observed in the acoustic branches: first, while the peaks corresponding to the out-of-plane and transverse modes (ZA and TA) are evident, the longitudinal mode (LA) is conspicuously missing. It has been theoretically proposed[49] that unlike transverse phonons, which are gauge phonons, longitudinal phonons cause unit cell volume changes that lead to charge fluctuations, whose screening should strongly suppress the EPC. However, screening should be relevant for momenta comparable to the Fermi momenta, while our measurement shows that the LA mode's absence extends to much higher momenta. Second, and even more surprising is the twist-angle dependence of the peak height (highlighted in color in a detailed view of the TA branches at positive and negative bias, Figs. 2f). Instead of decreasing with decreasing momentum, as typically anticipated for acoustic modes, the measured EPC actually increases.

To understand these unusual observations, we consider the relevant electron-phonon coupling mechanisms in TBG. Our experiment measures tunneling between the two layers, with electrons scattering between momentum states in their corresponding Dirac cones with a momentum shift of $q_M = 2K_D \sin\left(\frac{\theta}{2}\right)$ (Fig. 2b). This momentum mismatch is provided by a mini-BZ zone boundary phonon, commensurate with the moiré periodicity, emitted through electron-phonon coupling.

The simplest EPC occurs within the layer (an "in-layer" mechanism, Fig. 2g, SI 3): a phonon modulates the in-layer hopping amplitudes, $t_\parallel$, scattering the electron within the layer. However, since our measurement probes electrons tunnelling between the layers, this EPC appears in a second-order process: initially, an electron tunnels to a high-energy virtual state with the same momentum in the other layer, with amplitude $\alpha = \frac{t_\perp}{\hbar v_f K_D \theta} \ll 1$, followed by phonon emission through in-layer EPC, or vice versa (Fig. 2g). In TBG, there exists another "inter-layer" mechanism that couples electrons to anti-symmetric vibrations of the two layers (Fig. 2h), whose acoustic branch is the TBG's phason[19–24]. These anti-symmetric vibrations stretch the inter-layer bonds and therefore modify the tunnelling amplitudes between the layers, $t_\perp$ (Fig. 2h). This leads to an inter-layer EPC that already contributes to tunneling current in a first-order process (SI 3).

The unusual behavior of the phason's EPC becomes apparent in the limit of $q_M \to 0$. The strength of EPC is proportional to the stretching of the atomic bonds by a vibration whose amplitude is the phonon's zero-point-motion (ZPM), $\xi_{ZPM} = \sqrt{\hbar/2M\omega_M}$ ($\omega_M$ is the phonon frequency and $M$ is the carbon atom mass). For in-plane bonds, the stretching by $\xi_{ZPM}$ is distributed over many bonds, $N \sim (q_M a)^{-1}$ (Fig. 2g), and thus in the limit $q_M \to 0$, the corresponding in-layer EPC goes to zero, as expected for an acoustic mode. Interestingly, when the layers vibrate anti-symmetrically, the inter-layer bond stretching is directly given by $\xi_{ZPM}$, independent of $q_M$ (Fig. 2h). For a linearly dispersing phason ($\omega_M \sim q_M$) the zero point motion grows with decreasing $q_M$ as $\xi_{ZPM} \sim \frac{1}{\sqrt{q_M}}$, predicting that as long as $\hbar v_F q_M \gg t_\perp$ its inter-layer EPC should increase as $1/\sqrt{q_M}$ in the limit $q_M \to 0$. In contrast, if $q_M$ is held fixed and the wavevector $q \to 0$, the EPC to the phason mode vanishes[50]. This can be understood from the fact that a $q = 0$ displacement of the phason mode shifts the moiré pattern uniformly, which does not change the energy of the electronic states for incommensurate twist angles[50,51].

To study this quantitatively, we use the QTM's ability to tune and measure key experimental parameters in-situ. The inelastic conductance step in the experiment is directly related to the interlayer, $g_{inter-layer}$, and in-layer, $g_{in-layer}$, EPCs via (SI 3 and SI 4):

$$\Delta G = \beta\, A_{tip}\, \nu_B \nu_T (|g_{inter-layer}|^2 + \alpha^2 N_{layer} |g_{in-layer}|^2) \quad (1)$$

where $\beta = N_f N_b \frac{2\pi e^2}{\hbar} \frac{a^2 \sqrt{3}}{2}$, $N_f = 4$ is the flavor degeneracy (spin/valley), $N_b = 3$ is the number of Bragg scattering processes, $N_{layer} = 2$ is the number of layers, $a = 0.246 nm$ is lattice constant of graphene and $\alpha$ was defined above. The experiment-specific parameters in Eq. 1 are the tip contact area ($A_{tip}$) and the density of states (DOS) of the bottom and of the top layers ($\nu_B$ and $\nu_T$) all of which we can tune and measure in-situ.

To determine the dependence on tip contact area, we utilize unique QTM capabilities to both modify and image the tip area in-situ. The imaging is achieved by spatially scanning the tip along fixed atomic defects (on an adjacent area with defects in a WSe$_2$ layer, see SI 2). The measured current in such a spatial scan (Fig. 3b and 3c) reveals multiple copies of the tip shape, each produced when the tip overlaps a single atomic defect. After using mechanical manipulation away from the experimental region to change the tip size, imaging yields a tip area that is enhanced approximately two-fold (Fig. 3b, tip 1). Fig. 3a shows the $G$ vs $V_b$ traces measured for the two tip areas ($\theta = 16.8°$ in both), featuring conductance steps due to the ZA and TA phonon modes. When we plot $G$ normalized by the measured area, the curves collapse on each other (Fig. 3a inset), demonstrating that we capture the area dependence quantitatively.

To investigate the dependence on the DOS, we tune the carrier densities in both the top ($n_t$) and bottom ($n_b$) layers using a back gate, thereby tuning their DOS according to the Dirac relation: $\nu_T \sim \sqrt{|n_T|}$ and $\nu_B \sim \sqrt{|n_B|}$. Figs. 3d and 3e display $G$ and $d^2I/dV_b^2$ measured vs. back gate voltage, $V_{bg}$, and $V_b$, exhibiting steps in $G$ (peaks in $d^2I/dV_b^2$) that correspond to various phonon modes (labels). Notably, the amplitude of these steps progressively increases with $V_{bg}$. This is demonstrated in Fig. 3f, which plots $G$ vs $V_b$ along linecuts at $V_{bg} = 2V, 9V$ (dashed vertical lines in Fig. 3d). To relate these observations to the DOS, we employ an electrostatic model of the junction (SI 5) that accurately captures the neutrality points of both layers (red and blue dashed lines). The model shows that to a good approximation, both $n_b$ and $n_t$ vary linearly with $V_{bg}$, implying that $\nu_T \nu_B \sim \sqrt{|n_T||n_B|}$. Consequently, it predicts that $\Delta G$ should vary linearly with $V_{bg}$, and that the slope of this linear dependence should flips sign when the carriers' polarity changes. Fig. 3g plots the continuous evolution of $\Delta G_{TA}$ with $V_{bg}$ extracted from Fig. 3d, as well as traces extracted from similar maps measured at two other twist angles (SI 6). Interestingly, for all angles, the measurements show the predicted linear dependence of $\Delta G_{TA}$ on $V_{bg}$ and the slope's sign flipping upon carriers' polarity change (full model shown in dashed lines). Most strikingly, $\Delta G_{TA}$ increases with decreasing twist angle, showing that the coupling to the TA mode becomes stronger as $q_M$ decreases.

We can now determine the EPC from our experiments. Starting with the optical phonons - we extract the inelastic conductance step near $V_b = 160 mV$ from Fig. 2d, divide by 2 to account for the approximate degenerate LO and TO modes, normalize by the measured prefactors in Eq. 1, and plot the resulting quantity, $\Delta G/2\beta A_{tip}\nu_T\nu_B$, vs. $\theta$ in the inset of Fig. 4a (dots). This is compared with the theoretically predicted in-layer ($\alpha^2 N_{layer}|g_{in-layer}|^2$) and inter-layer ($|g_{inter-layer}|^2$) contributions, calculated using existing models (SI 3)[50]. The theory reveals that for optical phonons the inter-layer EPC is negligible, and thus we can directly extract the coupling $g_{TO}$ by identifying the measurement with the in-layer term. Plotted as a function of $\theta$ (main panel), we see that $g_{TO}$ is weakly dependent on $\theta$ and amounts to 300 – 350 meV, approximately twice larger than as determined by ARPES[25,26] and in good agreement with Raman[27,47].

Fig. 4c analyzes the EPC of the gauge (TA) acoustic phonon, comparing its measured and normalized conductance step ($\Delta G_{TA}/\beta A_{tip} \nu_T \nu_B$, plotted vs. $\theta$ in the inset (dots)) with theoretically calculated in-layer and inter-layer terms (lines). Notably, the in-layer coupling contributes negligibly to the gauge acoustic phonons EPC (SI 3) so that the inter-layer coupling, which exists only for a layer antisymmetric mode (phason mode) is dominant. Identifying the measured $\Delta G_{TA}$ with this mechanism thus directly provides the gauge phason EPC, $g_{phason}$, plotted vs. $\theta$ in the main panel. Remarkably, we see that $g_{phason}$ increases with decreasing twist angle as $1/\sqrt{q_M}$ (dashed line). The importance of this coupling becomes evident when it is compared with the measured mode energy, $\hbar\omega_{q_M}$, that diminishes linearly with $q_M$ (Fig. 4d). We note that the theory also explains the absence of the LA mode in our experiment (SI 3), resulting from an orthogonality between its momentum and polarization.

To contextualize our results to the observed superconductivity and strange metal behavior in MATBG, it is instructive to consider the dimensionless coupling, $\lambda = \frac{2\theta^2}{W} \frac{g_{q_M}^2}{\hbar\omega_{q_M}}$, obtained by integrating the EPC over the mini-BZ (SI 10, W is the flat band width). Our inelastic tunnelling measurements extend down to $\theta = 6°$, below which they are overwhelmed by elastic tunnelling current. However, to estimate the importance of EPC to the physics near the magic angle, we crudely extrapolate our results to lower angles, assuming that the observed dependence of the phason EPC on $q_M$ persists to the magic angle (see SI 10). We get that $\lambda_{gauge\_phason} = 1.1/W[meV]$. Additionally, we get that each optical mode contributes $\lambda_{optical} = 0.45/W[meV]$. Recalling that there are four intervalley-coupling optical modes (LO/TO, Top/Bottom layers), we see that the phason and optical modes have comparable and possibly important contributions to superconducting pairing. Optical phonons, on the other hand, have too high energy to contribute to the observed linear-in-T resistivity (strange metal) behavior, and here the coupling to the phason is the major source of scattering[24].

In summary, using the first realization of a cryogenic QTM, we demonstrate a novel technique for measuring phonon dispersions as well as mode- and momentum-resolved EPC in vdW materials. Our findings reveal that the phason coupling is at least as important to the physics of TBG as the coupling to optical phonons, and that both couple rather strongly to the electronic degrees of freedom. More generally, the method demonstrated here should be applicable to a broad class of vdW materials as well as to mapping the dispersions and couplings of other collective modes including plasmons[2], magnons[3] or spinons[4], making it a powerful new tool for studying collective behavior in quantum materials.

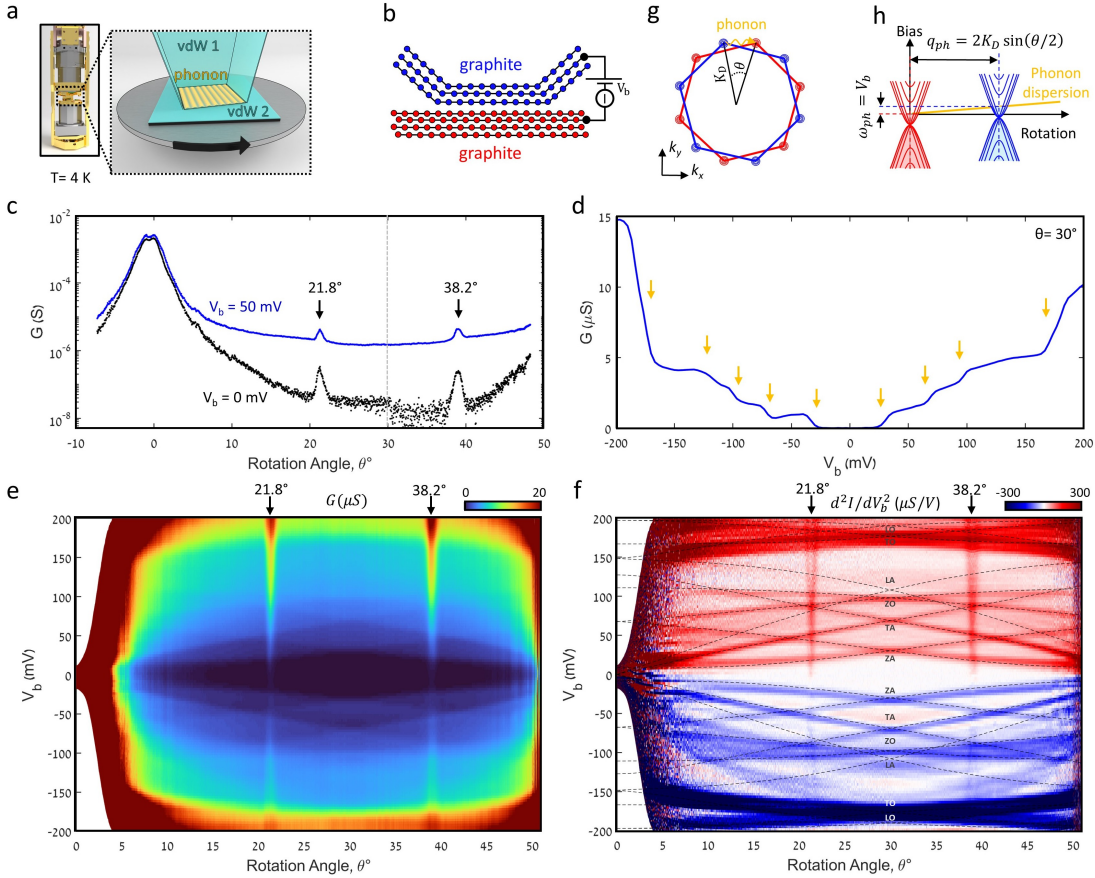

**Figure 1: Measuring phonon dispersion with the cryogenic QTM. a**. Schematics of the cryogenic QTM (inset), allowing us to form a continuously-twistable interface between two van der Waals materials at $T = 4K$ (main panel). We use an inelastic momentum-conserving electronic tunnelling process to emit and measure phonons at the interface with a well-defined momentum, continuously tunable by twisting (black arrow). **b**. The experiments in this figure are all performed in a twisted junction between two graphite flakes (few tens of nm thick), shown schematically. We apply a bias $V_b$ across the junction and measure current, $I$, and conductance $G = \frac{dI}{dV_b}$. **c**. Measured $G$ vs. twist angle, $\theta$, for $V_b = 0\ mV$ (black) and $50\ mV$ (blue). **d**. Measured $G$ vs. $V_b$ at $\theta = 30°$, exhibiting discrete steps (arrows), indicative of a series of inelastic tunnelling processes. **e**. Two-dimensional measurement of $G$ vs. $\theta$ and $V_b$, showing that the steps in $G$ appear at all twist angles and that their turn-on bias disperses smoothly with $\theta$. **f**. The second derivative, $d^2I/dV_b^2$, obtained numerically from panel **e**, highlighting the steps in $G$ that now appear as peaks. We see multiple peaks that disperse slowly with $\theta$, exhibiting a mirror symmetry around $\theta = 30°$, and showing good symmetry between positive and negative bias. In addition, we see sharply dispersing peaks near $\theta = 21.8°, 38.2°$. The theoretically calculated phonon spectrum of graphite[43] (dashed black) shows excellent agreement with the slowly dispersing peaks. From the measurement, we can identify the acoustic (TA, ZA) and optical (LO, TO, ZO) phonon branches. **g**. The Fermi surfaces in k-space of the top (blue) and bottom (red) graphite layers. **h**. The corresponding energy bands. At a finite twist angle, there is a momentum mismatch between these energy bands, and momentum-conserving electronic tunnelling between the layers can occur only via the emission of a phonon that provides the missing momentum, $q_{ph} = 2K_D \sin(\theta/2)$. This is an inelastic process that turns on when the bias voltage equals the phonon energy at this momentum $V_b = \omega_{ph}(q_{ph})$. By following the turn-on in the $\theta - V_b$ plane, we directly map the phonon dispersion line (orange).

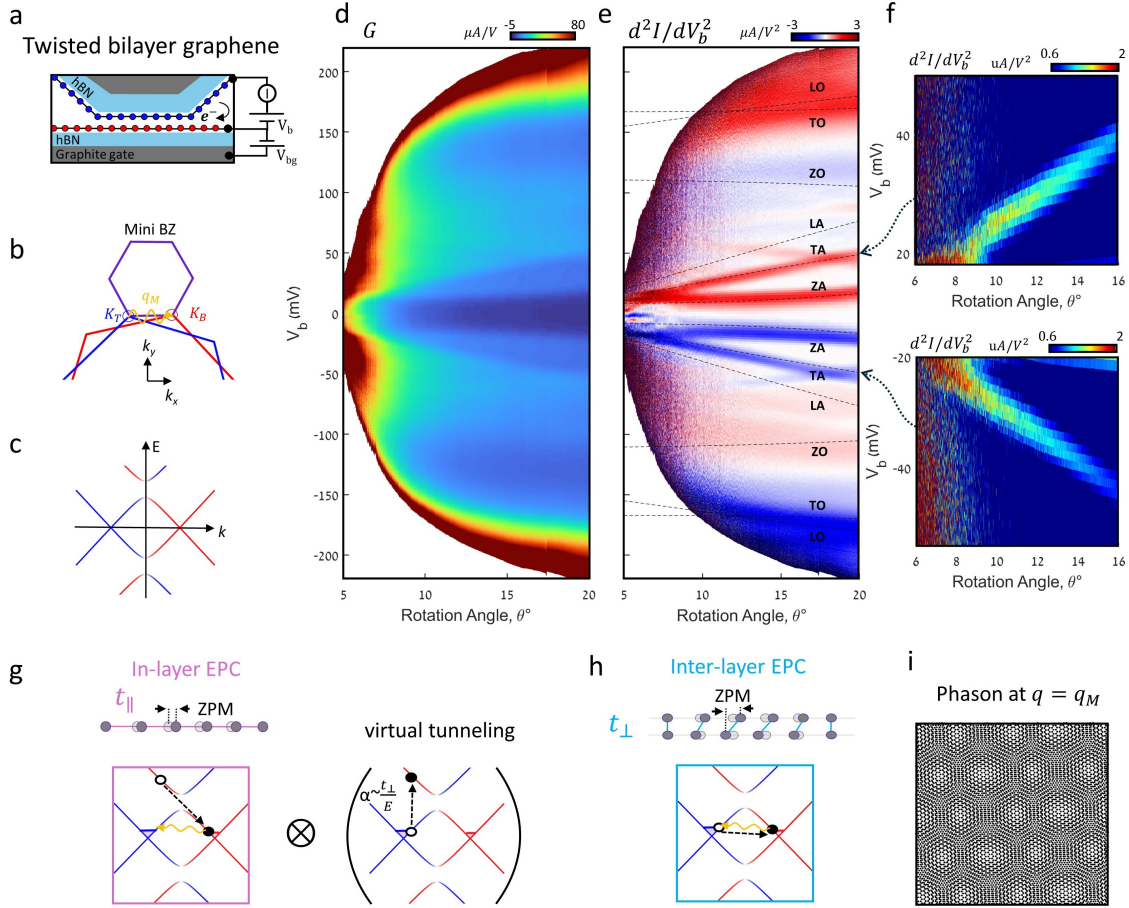

**Figure 2: Measured phonon spectrum and electron-phonon coupling (EPC) in Twisted Bilayer Graphene (TBG). a**. A tunable TBG is formed by bringing into contact a monolayer graphene on tip (blue) with a monolayer graphene on a flat sample (red), both backed by hBN, and with a global graphite back-gate. **b.** TBG's mini-Brillouin zone (BZ), with Dirac points of the top and bottom layers at its corners (labeled $K_T$ and $K_B$). The measured inelastic tunnelling processes involve an electron tunnelling between the layers while emitting a mini-BZ phonon that provides the missing momentum ($q_M$, orange). **c.** The corresponding TBG band structure along the $K_T - K_B$ line, colour-coded by layer weight. **d.** Measured conductance, $G$, vs. bias voltage, $V_b$, and twist angle, $\theta$, exhibiting steps in $G$ that disperse with $\theta$. **e.** The second derivative, $d^2I/dV_b^2$, obtained numerically from panel **d**, overlaid with the theoretically calculated phonon spectrum of graphite[43] (dashed lines, various modes labeled). **f.** Zoom-ins on the TA phonon peaks in panel **e**, at positive and negative bias, but with a colormap that reveals the variation of $d^2I/dV_b^2$ peak amplitude with $\theta$. **g.** In-layer EPC mechanism, originating from the phonon modifying the in-layer hopping amplitude, $t_\parallel$ (top illustration). EPC strength is proportional to the stretching of the bonds by a single phonon, namely with zero-point-motion amplitude (ZPM, marked). In the limit $q \to 0$, the ZPM stretching is distributed over many bonds and the EPC goes to zero. Moreover, since our experiments measure only electrons that tunnel between the layers, the in-layer EPC appears in them only as a second-order process – a virtual tunnelling between the layers (right brackets) followed by phonon emission (left square). **h**. Inter-layer EPC mechanism: here, an anti-symmetric motion of atoms in the two layers (a phason of the moiré lattice) directly modifies the tunnelling amplitude $t_\perp$ between them (top illustration). In this case, the maximal bond stretching is independent of $q_M$ and is equal to the full ZPM. Since the latter increases with decreasing $q_M$, counterintuitively, this EPC should intensify with decreasing $q_M$. Moreover, since here the phonon couples directly to the interlayer tunnelling, this process appears in first-order in our experiment (box). **i**. The experiment probes phasons with momentum $q = q_M$ (illustrated).

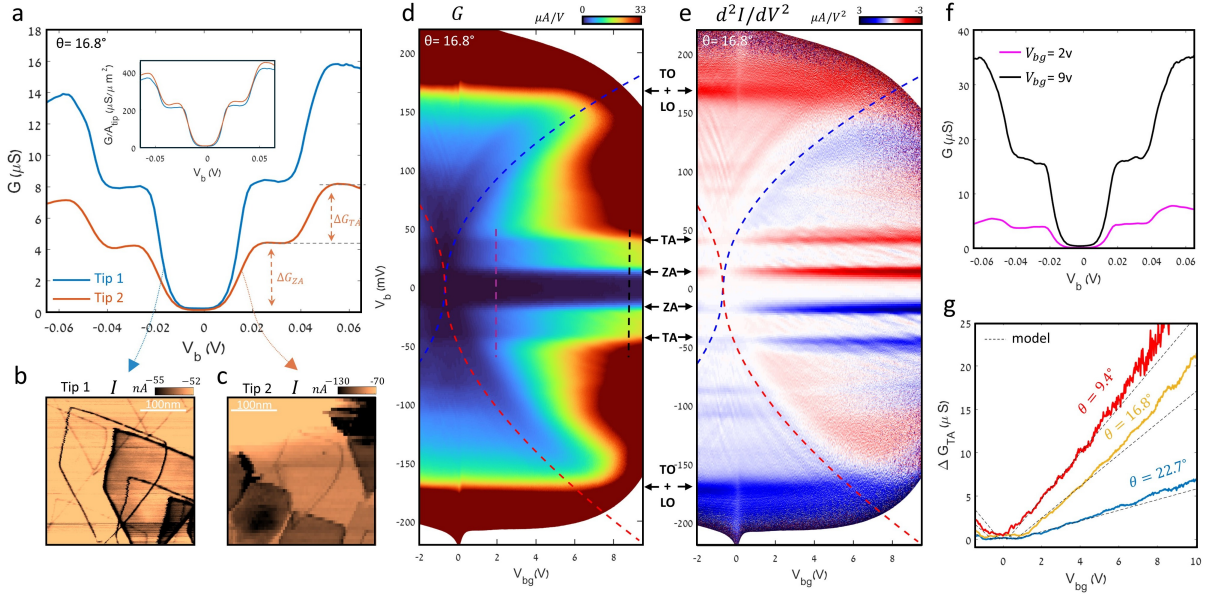

**Figure 3: Area and density of states (DOS) dependence of electron-phonon coupling induced inelastic tunnelling**. **a**. Tip area dependence of the inelastic tunnelling steps: measured $G$ vs. $V_b$ for TBG with $\theta = 16.8°$ and $V_{bg} = 4\,V$ for two tip contact areas, $A_{tip}$. The height of the inelastic steps corresponding to the ZA and TA phonon modes are marked ($\Delta G_{ZA}$, $\Delta G_{TA}$). Inset: same traces but normalized by the measured $A_{tip}$. **b.** and **c**. To image the tip area in-situ we scan it along fixed atomic defects located in a WSe$_2$ layer placed on top of the bottom graphene on the side of the main experiment. The panels show the imaged current in a spatial scan, exhibiting multiple copies of the tip shape, each produced by a different single defect. **d**. DOS dependence of the inelastic tunnelling steps: Measured G vs. $V_b$ and $V_{bg}$ for TBG with $\theta = 16.8°$. The gate voltage modifies the carrier density in the two layers and, consequently, their DOS. Dashed lines are the charge neutrality of the top (blue) and bottom (red) graphene layers, calculated from an electrostatic model of the junction (SI 5). Inelastic tunnelling steps in $G$ are visible as horizontal lines (corresponding phonon modes are marked). **e**. The second derivative, $d^2I/dV_b^2$, obtained numerically from panel **d**, exhibiting horizontal lines that correspond to the various phonon modes. We also see superimposed Fabry-Perot oscillations due to the finite size of the tip. **f**. Measured linecuts of $G$ vs. $V_b$ taken from panel d at $V_{bg} = 2\,V$ and $9\,V$ (vertical dashed lines in panel **d**). **g**. Amplitude of the conductance step of the TA mode, $\Delta G_{TA}$ vs. $V_{bg}$ extracted from panel **d** ($\theta = 16.8°$) and from two similar measurements in SI 6 done at $\theta = 9.4°$ and $22.7°$. Dashed lines plot the theoretical model that predicts a linear dependence of $\Delta G_{TA}$ on $V_{bg}$, nicely agreeing with the experiment. Strikingly, the overall $\Delta G_{TA}$ amplitude (and therefore the corresponding EPC) increases with decreasing $\theta$.

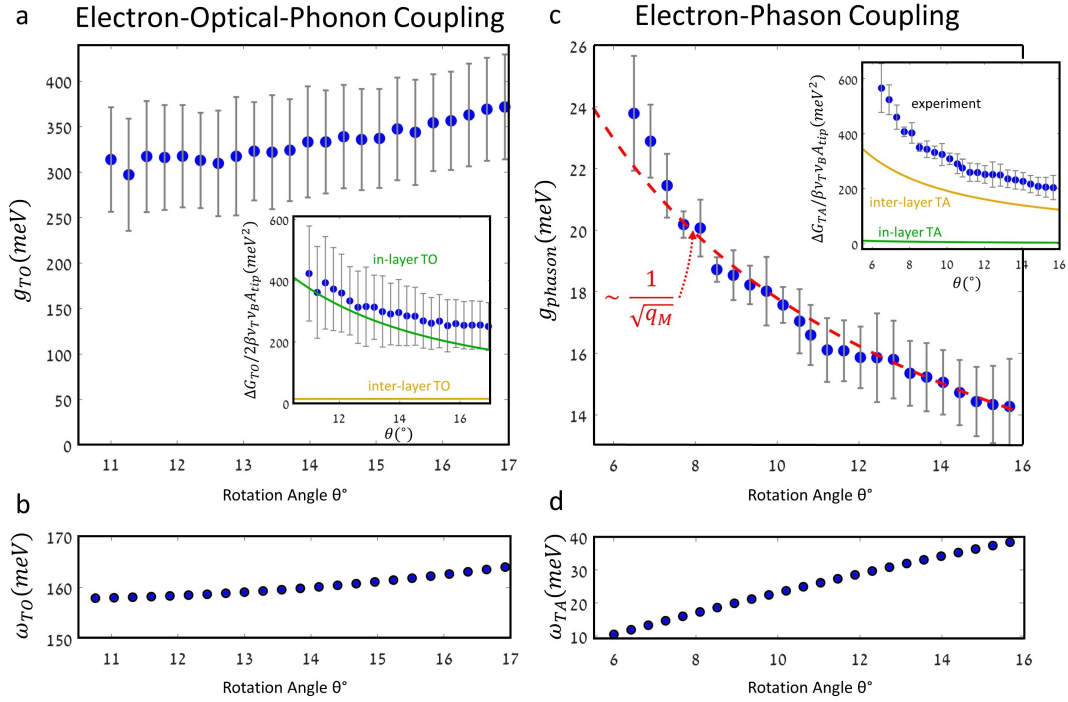

**Fig 4: Electron-optical-phonon and electron-phason coupling in TBG. a.** Inset: Measured inelastic conductance step, vs. $\theta$, corresponding to the intervalley optical TO mode, normalized following Eq. 1 using the measured tip area and densities of states (we divided the measured step by 2 to account for the approximate degenerate TO and LO modes). Solid lines are theoretically calculated in-layer (green) and inter-layer (yellow) contributions. Visibly, for optical phonons the dominant EPC mechanism is the in-layer one (SI 3). Main panel: The intervalley optical TO mode EPC, $g_{TO}$, determined from the measurements in the inset. Error bars are obtained from differences between measurements at positive and negative bias and all other experimental uncertainties. **b.** TO phonon energy as a function of $\theta$ extracted from Fig. 2e. **c.** Inset: Measured inelastic conductance step, vs. $\theta$, corresponding to the acoustic TA mode, normalized following Eq. 1 using the measured tip area and densities of states. Solid lines are the theoretically calculated in-layer (green) and inter-layer (yellow) contributions. In contrast to optical phonons, the acoustic phonons' EPC is dominated by the inter-layer coupling mechanism, which is the layer-antisymmetric phason mode. Main panel: The electron-phason coupling, $g_{phason}$, determined from the measurements in the inset. Error bars are obtained from differences between measurements at positive and negative bias and all other experimental uncertainties. **d**. TA mode energy as a function of $\theta$ extracted from Fig. 2e.

**Methods:**

*Cryogenic (T=4K) quantum twisting microscope*: The cryogenic QTM consists of two nanopositioners towers facing each other. One tower is equipped with three translational degrees of freedom (XYZ) and carries a flat sample. The opposing tower holds an atomic force microscope cantilever (AFM) and has a rotational degree of freedom (θ) as well as two lateral degrees of freedom for positioning the AFM tip at the center of rotation. The entire assembly is cooled down in vacuum to liquid helium temperatures. In the experiment we bring into contact a van der Waals (vdW) heterostructure on the tip with a flat vdW heterostructure on the bottom sample, creating a two-dimensional interface, typically a few hundred nanometers across in both directions. vdW attraction between the two heterostructures self-cleans the contact area, leading to a pristine interface. The tip and sample remain in continuous contact throughout the experiment, including during any twisting or scanning operation. Self-sensing AFM cantilevers equipped with piezoresistive elements are used for monitoring and maintaining a constant force, ensuring that the contact interface's area remains constant during the scans. The QTM junctions in the current paper do not have a tunnelling barrier separating the two conducting sides (apart from Fig. 3b and 3c where we used defects in $WSe_2$ barrier to image the tip shape). This means that the resistance of the tunnel junction, especially at low twist angles or at high bias can be comparable or lower than the resistance of the contacts or of the bulk of the 2D layers leading from the contact to the junction. We obtain this contact resistance from measurement close to zero degrees twist angle, and then directly calculate how much bias drops on the contact resistance and how much is dropping on the junction, $V_b$, which is the variable we use throughout the paper (SI 7).

**Devices:**

*Fabrication of the vdW-devices-on-tip*: The fabrication of the vdW-devices-on-tip follows the procedure we described previously[1]. In brief, we use tipless AFM cantilevers that have a thin metallic line reaching its end, which we use for electrical connecting to the active vdW layer. We make a custom tip using focused ion beam (FIB) deposited platinum pyramid, with base dimensions of 2x2 um and a height of 1-2 um. The vdW layers are transferred onto the pyramid using the polymer membrane transfer technique.

*Conductance measurements*: Voltages are applied using a custom-built DAC consisting of a.c. + d.c. sources. Current are measured using a Femto amplifier, followed by an NI sampler. We use a Zurich Instruments lock-in amplifier (MFLI) for force feedback of the piezoresistive AFM cantilevers.

**Acknowledgements:** We thank Ado Jorio, Efthimios Kaxiras, Francisco Mauri, Stephanie Reich and Jonathan Ruhman for fruitful discussions. Work was supported by the Leona M. and Harry B. Helmsley Charitable Trust grant, the Rosa and Emilio Segre Research Award, the ERC-Adg grant (QTM, no. 101097125), the DFG funded project Number 277101999 - CRC 183 (C02) and the BSF grant (2020260). A.I was supported by the Azrieli Fellow Program. E.B was supported by the European Research Council (ERC) under grant HQMAT (Grant Agreement No. 817799) and CRC 183 of the Deutsche Forschungsgemeinschaft (Project C02). L.G acknowledges the support by NSF Grant No. DMR-2410182 and by the Office of Naval Research (ONR) under Award No. N00014-22-1-2764. F.G acknowledges support from the Severo Ochoa programme for centres of excellence in R&D (CEX2020-001039-S / AEI / 10.13039/501100011033, Ministerio de Ciencia e Innovacion, Spain) from the grant (MAD2D-CM)-MRR MATERIALES AVANZADOS-IMDEA-NC, NOVMOMAT, Grant PID2022-142162NB-I00 funded by MCIN/AEI/ 10.13039/ 501100011033 and by "ERDF A way of making Europe". F.v.O acknowledges the support by the Deutsche Forschungsgemeinschaft through CRC 183 (project C02 and a Mercator fellowship) and a joint ANR-DFG project (TWISTGRAPH).

**Author Contributions:** J.B, J.X, A.I. and S.I. designed the experiment., J.B, J.X and A.I. built the cryogenic QTM microscope, fabricated the devices, and performed the experiments. J.B, J.X, A.I. and S.I. analyzed the data. J.X, E.B., L.G., F.G. and F. v. O. wrote the theoretical models K.W. and T.T. supplied the hBN crystals. J.B, J.X, A.I. and S.I. wrote the manuscript with input from other authors.

**Data availability:** The data shown in this paper is provided with this paper. Additional data that support the plots and other analysis in this work are available from the corresponding author upon request.

**Competing interests:** The authors declare no competing interests.

**Correspondence and requests for materials** should be addressed to shahal.ilani@weizmann.ac.il